\documentclass[12pt,english]{article}
\usepackage{bm,bbm,lscape}
\usepackage[T1]{fontenc}
\usepackage{natbib,amssymb,amsmath,authblk}
\usepackage{marvosym}
\usepackage{multicol}
\usepackage[british]{babel}
\usepackage{graphicx,xcolor,geometry,setspace,lineno}
\usepackage{pstricks}
\usepackage[british]{babel}
\usepackage[justification=justified,singlelinecheck=false,labelfont=it]{caption}
\usepackage{graphicx}
\usepackage{fancyhdr}
\definecolor{Mygrey}{gray}{0.65}

\title{Maximum penalized likelihood estimation in semiparametric mark-recapture-recovery models}
 
\author{
  Th\'eo Michelot\\ \vspace{-1em}
  \textit{INSA de Rouen}\\ \vspace{1em}
  Roland Langrock\\ \vspace{-0.2em}
  \textit{University of St Andrews}\\  \vspace{1em}
  Thomas Kneib\\ \vspace{-0.2em}
  \textit{Georg August University G\"ottingen}\\  \vspace{1em}
  Ruth King\\ \vspace{-0.2em}
  \textit{University of St Andrews} 
}

\date{}

\begin{document}

\begin{spacing}{1.1}
\maketitle


\begin{abstract}
We discuss the semiparametric modeling of mark-recapture-recovery data where the temporal and/or individual variation of model parameters is explained via covariates. Typically, in such analyses a fixed (or mixed) effects parametric model is specified for the relationship between the model parameters and the covariates of interest. In this paper, we discuss the modeling of the relationship via the use of penalized splines, to allow for considerably more flexible functional forms. Corresponding models can be fitted via numerical maximum penalized likelihood estimation, employing cross-validation to choose the smoothing parameters in a data-driven way. Our contribution builds on and extends the existing literature, providing a unified inferential framework for semiparametric mark-recapture-recovery models for open populations, where the interest typically lies in the estimation of survival probabilities.
The approach is applied to two real datasets, corresponding to grey herons ({\it Ardea Cinerea}), where we model the survival probability as a function of environmental condition (a time-varying global covariate), and Soay sheep ({\it Ovis Aries}), where we model the survival probability as a function of individual weight (a time-varying individual-specific covariate). The proposed semiparametric approach is compared to a standard parametric (logistic) regression and new interesting underlying dynamics are observed in both cases.
\end{abstract}

\vspace{0.5em}

\section{Introduction}\label{intro}

Capture-recapture studies are commonly conducted on wildlife populations in order to better understand the underlying population dynamics. For example, interest may be focused on estimating survival probabilities. Within these studies, researchers go into the field in a series of capture events. Individuals are uniquely marked at their initial capture, for example using a tag/ring, or by recording unique natural markings. At each subsequent survey, all individuals observed are recorded, and those that have not been observed previously are again uniquely identified, before all are released back into the population. We assume that individuals can be observed alive or recovered dead in each survey. The resulting data can be summarized as the observed encounter histories for all individuals observed within the population, detailing for each survey event whether an individual was not observed, observed alive or recovered dead. We consider Cormack-Jolly-Seber-type models, which are models for open populations where the interest typically lies in the estimation of survival probabilities (see \citealp{schs00}, and \citealp{kin14}, for a review of these models). The original Cormack-Jolly-Seber (CJS) model considered only live captures (i.e.\ capture-recapture data), and was extended to additional recoveries by \citet{bar97} and \citet{cat98}. The likelihood of such models is a function of survival, recapture and recovery probabilities.

In the absence of additional covariate information, latent class models \citep{plepn03} and random effect models \citep{gimc10,roy08} have been fitted to CJS models to incorporate unobserved heterogeneity. However, in the presence of additional covariate information, the model parameters are often expressed as an explicit function of these underlying factors in order to explain individual, temporal and/or spatial heterogeneity. Possible covariates include, for example, individual-level factors (e.g.\ condition, gender) and (global) environmental-level factors (e.g.\ resource availability, climatic condition). Typically, a relationship is specified in the form of a (parametric) generalised linear model between the model parameters and the given covariate(s) (though \citealp{barf06} also specify a relationship between individual covariates and the latent class probabilities for individuals). For example, for the survival probabilities a logistic regression with linear or quadratic predictors can be used \citep{mor13}. The parametric relationship specified between the model parameters and the covariates is often assumed with little investigation, if any, into the absolute or relative goodness of fit.

\citet{gim06} were the first to use splines in a semiparametric modeling approach to flexibly model the relationship between the model parameters and the covariate of interest. Such nonparametric approaches allow for more flexible, data-driven functional relationships.
\citet{gim06}, as well as \citet{bonts09} and \citet{gim09}, fit the semiparametric models in a Bayesian framework. Frequentist semiparametric analyses of mark-recapture-recovery data are presented by \citet{few09}, \citet{via10} and \citet{hug12,hug12b}. However, all these analyses focus on specific types of covariates, and none of the existing frequentist approaches addresses the important case of individual-specific covariates that evolve stochastically over time (but note that \citealp{via10}, considers an individual-specific covariate that is a deterministic function of time, namely age). In such a scenario a well-known difficulty arises, namely that when an individual is unobserved, the corresponding covariate value is also unknown, leading to a (typically) significant proportion of unknown (or missing) covariate values. The special case of discrete covariates leads to the Arnason-Schwarz model \citep{schwarz93,brownie93}, where the likelihood is available in closed form. In the case of continuous covariates, there is a substantial body of literature dealing with the estimation of fully parametric CJS-type models in the presence of missing covariate data. Existing approaches include the use of a conditional (``trinomial'') approach, deriving a likelihood conditional on only the observed covariate values \citep{catmt08}, a multiple imputation approach \citep{workb14}, a Bayesian data augmentation approach \citep{bons06,kinbc08,kinmgb09} and a numerical integration approach leading to an approximate likelihood that can be evaluated efficiently using the hidden Markov model machinery \citep{lan13}.

Here we aim to build on the existing literature by providing a frequentist inferential framework for semiparametric mark-recapture-recovery models that allows for the consideration of all the different types of covariates -- environmental, time-constant individual and time-varying individual covariates -- using a unified machinery. In contrast, previous semiparametric approaches focused exclusively on environmental covariates \citep{gim06,hug12}, individual-specific but time-constant or deterministic covariates \citep{via10,hug12b}, or the (most challenging) case of individual-specific and stochastically time-varying continuous covariates. The latter has been considered in \citet{bonts09} using Bayesian inference for adaptive spline models within a reversible jump Markov chain Monte Carlo framework.
We provide a frequentist framework which unifies previous semiparametric capture-recapture-recovery modelling approaches for any form of continuous covariate information (time- and/or individual-constant/varying). The two key components of our approach are: (i) the (non-penalized) likelihood for the CJS-type model under consideration (with the structure of the likelihood depending on the given type of covariate) and (ii) the penalty term imposed on nonsmoothness of the spline-based functional estimate of the relationship between the parameter(s) of interest and the covariate. Smoothing parameters that control the balance between goodness-of-fit and smoothness are chosen either via cross-validation or via an information-criterion-based score, and in each case from a pre-specified set of candidate values. Since parametric polynomial models are included as a limiting case for very large smoothing parameters, both approaches allow for the functional forms to effectively reduce to the parametric limiting cases.  %

The feasibility and the relevance of the presented techniques are demonstrated in two real data applications. The first corresponds to the case of environmental covariates, investigating the relationship between survival probabilities of grey herons (\textit{Ardea Cinerea}) and winter weather condition. In the UK, the grey heron has been extensively monitored by the British Trust of Ornithology (BTO), in particular since it represents an important indicator species due to its status as a top predator in freshwater ecosystems \citep{mar04}. The second application corresponds to the more challenging case of individual-specific and time-varying covariates, looking at how individual body mass affects the survival probability of Soay sheep ({\it Ovis Aries}). These sheep have been the subject of numerous studies on population dynamics, due to their isolated nature with no natural predators and the ease with which individuals can be marked and recaptured. 

In Section \ref{model}, we introduce the notation and describe the likelihoods of CJS models, in particular for the different types of covariates, before describing the spline-based semiparametric modeling approach. In Section \ref{infer}, we provide details on the inferential framework, including uncertainty quantification and the strategies for choosing the smoothing parameter(s). In Section \ref{simul}, we conduct a simulation study to investigate the performance of the method in the most challenging case with individual-specific and time-varying covariates. Finally, in Section \ref{appl}, we apply the approach to the two different sets of real data.

\section{Semiparametric mark-recapture-recovery models}\label{model}

In this section, we formulate the semiparametric modeling approach for different types of covariates. After introducing relevant notation in Section \ref{not}, we briefly describe the standard CJS model in Section \ref{cjs}. We then review the three possible types of covariates, in each case providing the (non-penalized) likelihood (Sections \ref{envi}--\ref{inditv}) of corresponding models. Finally, in Section \ref{npreg}, we discuss how to use the likelihood in concert with P-splines in order to implement a semiparametric approach in all three cases.

\subsection{Data and notation}\label{not}

Suppose that there are $T$ capture occasions and $N$ observed individuals within the study period. The capture history for individual $i=1,\dots,N$ is $\{x_{i,t}:t=1,\dots,T\}$, such that
 \begin{linenomath*}
\begin{equation*}
x_{i,t} = \left\{ \begin{array}{ll}
1 & \mbox{ if individual $i$ is observed at time $t$; } \\
2 & \mbox{ if individual $i$ is recovered dead in the interval $(t-1, t]$; } \\
0 & \mbox{ otherwise. }
 \end{array} \right.
\end{equation*}
 \end{linenomath*}
Note that for a given time $t$, we make the distinction between ``recently dead'' individuals (death in $(t-1,t]$) and ``long dead'' individuals (death before $t-1$), and assume that only recently dead individuals can be recovered dead at a given capture event. This is a standard assumption for mark-recapture-recovery data, due to the decay of marks for identifying individuals once they have died.
The extension to the case where it is possible for ``long dead'' individuals to be recovered is straightforward (see for example \citealp{catfmn01}, and \citealp{kinl15}, for further discussion).

We consider CJS models where the likelihood is a function of survival, recapture and potentially also recovery probabilities. For a given observed individual $i=1,\dots,N$ and discrete capture occasions $t=1,2,\ldots,T$, we use the notation:
 \begin{linenomath*}
\begin{align*}
\phi_{i,t} &= \Pr \bigl( \text{individual $i$ alive at time } t+1 \, \vert \, \text{individual $i$ alive at time } t \bigr)  , \\
p_{i,t} &=  \Pr \bigl(  \text{individual $i$ observed at time } t \, \vert \, \text{individual $i$ alive at time } t \bigr)  , \\
\lambda_{i,t} &=  \Pr \bigl(  \text{individual $i$ recovered at time } t \, \vert \, \text{individual $i$ died in } (t-1,t] \bigr)  .
\end{align*}
 \end{linenomath*}
For notational convenience, in what follows we drop the subscripts on individual $i$ when the parameters are common to all individuals at time $t$.

\subsection{CJS model}\label{cjs}

We initially consider the standard CJS model when there are no covariates present. For this model, we condition on the first time an individual is observed within the study. The data can be summarized in the form of m-arrays and/or d-arrays, which constitute sufficient statistics \citep{mor13} and permit efficient likelihood evaluation. The m-array contains the information on live recaptures (and on the number of individuals not observed again after a given capture occasion), and the d-array contains the information on dead recoveries. For $r=1,\ldots,T-1$, $s=r+1,\ldots,T$, let $m_{rs}$ denote the number of individuals released at time $r$ and next observed at time $s$. For $r=1,\ldots,T-1$, let $m_{r,T+1}$ denote the number of individuals released at time $r$ and not observed again. Similarly, for $r=1,\ldots,T-1$ and $s=r+1,\ldots,T$, let $d_{rs}$ denote the number of individuals that were released at time $r$ and next observed dead within the interval $(s-1,s]$. For $r=1,\ldots,T-1$, the cell probabilities for the m-array are given by $q_{rs}^{(m)}$, such that,
 \begin{linenomath*}
\begin{equation*}
q_{rs}^{(m)}=
\begin{cases}
\left[ \prod_{t=r}^{s-2} \phi_t (1-p_{t+1})  \right] \phi_{s-1} p_{s} & \quad \text{for } r<s<T+1; \\
\chi_r & \quad \text{for } s=T+1,
\end{cases}
\end{equation*}
 \end{linenomath*}
where we use the convention that the empty product is equal to 1 (for $s=r+1$), and where $\chi_r$ denotes the probability an individual is not observed after time $r$ (either alive or dead). These probabilities can be calculated using the recursion
 \begin{linenomath*}
\begin{equation*}
 1-\chi_{r} = (1-\phi_{r}) \lambda_{r+1} + \phi_{r}\bigl(1-(1-p_{r+1})\chi_{r+1}\bigr),
\end{equation*}
  \end{linenomath*}
for $r=1,\ldots,T-1$, determined recursively starting with $\chi_{T}=1$. The left hand side of this equation corresponds to the probability of being observed after time $r$; the right hand side is the sum of the probabilities associated with dying and being recovered dead in the interval $(r,r+1]$ and surviving until time $r+1$ and being observed again. For further explanation of the calculation of this recursion, see for example \cite{cat98,mor13}. Note that in this model the recapture, recovery and survival probabilities are assumed to be common to all individuals at any time $t$, so that we omit the subscript $i$.

Similarly, for $r=1,\ldots,T-1$ and $s=r+1,\ldots,T$, the cell probabilities for the d-array are given by $q_{rs}^{(d)}$, where
 \begin{linenomath*}
\begin{equation*} q_{rs}^{(d)} = \left[ \prod_{t=r}^{s-2}\phi_t(1-p_{t+1}) \right] (1-\phi_{s-1})\lambda_{s} \, .
\end{equation*}
 \end{linenomath*}

The corresponding likelihood is of multinomial form:
 \begin{linenomath*}
\begin{equation} \label{lik1}
\mathcal{L} \propto \prod_{r=1}^{T-1} \left[ \prod_{s=r+1}^{T+1} \left( q_{rs}^{(m)} \right)^{m_{rs}} \prod_{s=r+1}^T \left( q_{rs}^{(d)} \right)^{d_{rs}} \right] \,.
\end{equation}
 \end{linenomath*}
See for example, \cite{mor13} for further discussion of the likelihood and associated assumptions for the CJS model.

We now extend the CJS model where the model parameters are a function of covariates. We distinguish three types of covariates: time-varying global covariates, time-constant individual-specific covariates and time-varying individual-specific covariates. 

\subsection{CJS model with time-varying global covariates}\label{envi}

We consider the case of time-varying global covariates, i.e.\ covariates that vary over capture occasions but are not individual-specific. For example, such a covariate could correspond to the climatic condition at different times. The model parameters are again only a function of time, as it is assumed that the covariates are common to all individuals (so that the subscript $i$ on the parameters can be dropped). Assuming that all the time-varying covariates are observed, the model parameters are typically expressed as a deterministic function of the covariates. Thus, the corresponding likelihood is the same as in Equation (\ref{lik1}), allowing for the covariate dependence on the parameters. We note that it is this likelihood that we consider in Section \ref{herons} below.

\subsection{CJS model with time-constant individual-specific covariates}\label{indi}


In the case where parameters of the CJS model are driven by time-constant or deterministic individual-specific covariates, we need to separately evaluate the probability of each encounter history, since even two identical histories will in general have different probabilities due to the individual-specific covariates. For individual $i$, let $c_i$ denote the time of its initial (live) capture, and let $l_i$ the occasion on which it is last known to be alive (corresponding to the final occasion an individual is observed alive and not seen again, or to the capture occasion just before an animal is recovered dead).

The likelihood for the mark-recapture-recovery data is the product over each individual of the
corresponding probability of their observed encounter history:
 \begin{linenomath*}
\begin{align*}
\mathcal{L} & =  \prod_{i=1}^N \biggl[ \bigl( (1-\phi_{i,l_i}) \lambda_{i,l_i}\bigr)^{I_{\{ x_{i,l_i+1}=2\}}}\chi_{i,l_i}^{1-I_{\{ x_{i,l_i+1}=2\}}} \\
            & \quad\quad \times \prod_{r=c_i}^{l_i-1} \phi_{i,r} p_{i,r+1}^{I_{\{ x_{i,r+1}=1\}}}(1-p_{i,r+1})^{1-I_{\{ x_{i,r+1}=1\}}} \biggr] ,
\end{align*}
 \end{linenomath*}
where $I$ denotes the indicator function; see, for example, \citet{cat98,catmt08}. The first term in the first line corresponds to individual $i$ being recovered dead at time $l_i+1$, the second term in the first line to individual $i$ not being observed after time $l_i$ (when not recovered dead), and the second line to the capture history from time $c_i$ to time $l_i$ (when it is known that individual $i$ is alive), conditional on being initially observed at time $c_i$.

In the case of individual-specific covariates that do not vary stochastically over time, it is this likelihood that needs to be considered. The survival, recovery and recapture probabilities may be time-independent, in which case the corresponding subscript can be omitted. However, in some cases, for example with a covariate indicating the (known) age of individuals (as in \citealp{via10}) or with year-dependent recapture probabilities, these quantities may be time-dependent. The important case of stochastically varying individual-specific covariates requires a separate consideration, given in Section \ref{inditv} below, due to the missing data arising in such a scenario.

\subsection{CJS model with time-varying individual-specific covariates}\label{inditv}

We now turn our attention to the most challenging case where covariates are individual-specific and time-varying. In the considered scenario, we regard each encounter history as the outcome of a state-space model, where the (partially) hidden system process involves the survival states and the covariate values for the corresponding individual. A detailed description of the corresponding model fitting strategy is given in \citet{lan13}, and in the following we only summarize the main steps. For individual $i$, the survival process, $(s_{i,c_i},\ldots,s_{i,T})$, is defined such that
 \begin{linenomath*}
\begin{equation*}
s_{i,t} = \begin{cases} 1 \;\;\; \text{if individual $i$ is alive at time } t; \\
2 \;\;\; \text{if   individual  $i$  is dead at time } t, \text{ but was alive at time } t-1; \\
3 \;\;\; \text{if  individual $i$ is dead at time } t, \text{ and was dead at time } t-1. \end{cases}
\end{equation*}
 \end{linenomath*}
We note that the survival states for individual $i$ (following initial capture) are fully known if they are observed at the final capture time, i.e.\ if $l_i = T$. Similarly, the survival states are fully known if individual $i$ is recovered dead within the study, i.e.\ if $x_{i,l_i+1} = 2$, then $s_{i,t} = 1$ for $t=c_i,\dots,l_i$; $s_{i,l_i+1} = 2$ and $s_{i,t} = 3$ for $t=l_i+2,\dots,T$ (for $l_i \le T-2$). However, for an individual that is not observed at time $T$ or recovered dead within the study, their survival state is unknown following their final sighting, i.e.\ $s_{i,t} = 1$ for $t=c_i,\dots,l_i$ but $s_{i,t}$ is unknown for $t=l_i+1,\dots,T$. Notationally, let $\mathcal{S}_i^c$ denote the set of all times at which the survival state of individual $i$ is unknown, and let $\mathcal{W}_i^c$ denote the set of all times at which the covariate value for individual $i$ is unknown, following their initial capture. We note that we consider the general case here where the covariate value may not always be recorded when an individual is observed within the study.
Assuming a first-order Markov process for the covariate process -- specified by an initial distribution with density $f_0$ and the conditional distribution $f(w_{i,t}|w_{i,t-1})$ -- the likelihood of the encounter histories and the observed covariate values, conditional on the initial capture events, can be written in the form
 \begin{linenomath*}
\begin{align*}
 \mathcal{L}   = \prod_{i=1}^N \biggl[ & \int \ldots  \int \sum_{\tau \in \mathcal{S}_i^c} \sum_{s_{i,\tau} \in \{ 1,2,3\}}  f_0(w_{i,c_i}) \\ \nonumber  & \times \prod_{t=c_i+1}^T f(s_{i,t}\vert s_{i,t-1},w_{i,t-1})  f(x_{i,t}\vert s_{i,t},w_{i,t})f(w_{i,t}\vert w_{i,t-1}) d\mathbf{w}_{i,\mathcal{W}_i^c} \biggr] .
\end{align*}
 \end{linenomath*}
Here $f(s_{i,t}\vert s_{i,t-1},w_{i,t-1})$ corresponds to the survival process, $f(x_{i,t}\vert s_{i,t},w_{i,t})$ to the recapture and recovery processes, and $f(w_{i,t}\vert w_{i,t-1})$ to the evolution of the covariate value according to the assumed Markov process. In general, for continuous covariates the necessary integration within this likelihood expression is analytically intractable. However, a discretization of the range of possible covariate values -- considering a partition of the range into $m$ intervals -- leads to a closed-form expression for an approximate likelihood, where the approximation can be made arbitrarily accurate by increasing $m$. The approximation to the likelihood becomes extremely accurate already for values of $m$ around 20-40 \citep{lan13}. Such an approach is feasible in typical mark-recapture-recovery studies since the approximate likelihood can be evaluated using the efficient hidden Markov model machinery, and in particular the forward algorithm. In the case of discrete covariates the Arnason-Schwarz model can be used, where the (exact) likelihood is available in closed form \citep{schwarz93}.

\subsection{Nonparametric regression in mark- \\ recapture-recovery models}\label{npreg}

We consider scenarios in which some of the quantities of interest -- such as the survival, recapture and recovery probabilities -- are modeled as deterministic functions of a covariate $w_{i,t}$, which may be of any of the types described in Sections \ref{envi}--\ref{inditv}. We focus on the case of a single covariate, noting that there is no in-principle problem to applying the suggested methodology also in the case of multiple covariates (see the discussion in Section \ref{discuss}). The covariate may be individual-specific and/or time-varying, hence the indices $i$ and $t$. 
In the mark-recapture-recovery setting, one can for example use the logistic function to link the survival probability to the covariate $w_{i,t}$. In a typical parametric model,
 \begin{linenomath*}
\begin{equation}\label{linreg}
	g(\phi_{i,t})=\text{logit}(\phi_{i,t}) = \beta_0+\beta_1 w_{i,t},
\end{equation}
 \end{linenomath*}
where $\beta_0$ and $\beta_1$ are regression parameters to be estimated; see, for example, \cite{nor79}. In the following, we focus on the link between survival probability and covariate, noting that other quantities of interest, such as the recapture and recovery probabilities, can be modeled analogously, and that in general any appropriate link function $g$ can be considered. In order to increase the flexibility of the functional form, we express the predictor in (\ref{linreg}) as a finite linear combination of basis functions, $B_{1}, \ldots , B_{K}$, each of them evaluated at $w_{i,t}$:
 \begin{linenomath*}
\begin{equation}\label{lincom}
	\text{logit}(\phi_{i,t}) = \sum_{k=1}^K \gamma_k B_k(w_{i,t}),
\end{equation}
 \end{linenomath*}
where $\gamma_1,\dots,\gamma_K$ are coefficients to be estimated. In principle, any functions $B_{1},\ldots,B_K$ can be used as a basis. A common choice is to use B-splines, which form a numerically stable, convenient basis for the space of polynomial splines, i.e.\ piecewise polynomials that are fused together smoothly at the interval boundaries; see \citet{deb78} and \citet{eil96} for more details. Throughout this manuscript, we use cubic B-splines  with equidistantly spaced knots, in ascending order in the basis used in (\ref{lincom}). The number of B-splines considered, $K$, determines the flexibility of the functional form, as an increasing number of basis functions allows for an increasing curvature of the predictor. Instead of trying to select an optimal number of basis elements, we follow \citet{eil96} and modify the likelihood by including a difference penalty on coefficients of adjacent B-splines. The number of basis B-splines, $K$, needs to be sufficiently large in order to reflect the structure of the estimated function. Once this threshold is reached, a further increase in the number of basis elements no longer changes the fit to the data due to the impact of the penalty. Considering second-order differences -- which leads to an approximation of the integrated squared curvature of the predictor \citep{eil96} -- leads to the difference penalty
 \begin{linenomath*}
\begin{equation*}
 \mathcal{P}(h) = \dfrac{h}{2} \sum_{k=3}^K (\Delta^2 \gamma_k)^2\ ,
\end{equation*}
 \end{linenomath*}
where $h \geq 0$, $\Delta^2 \gamma_k = \gamma_k-2\gamma_{k-1}+\gamma_{k-2}$. In each of the CJS-type models considered in Sections \ref{envi}--\ref{inditv}, we can simply subtract $\mathcal{P}(h)$ from the corresponding log-likelihood of the model, yielding a penalized log-likelihood. The maximum penalized likelihood estimate then reflects a compromise between goodness-of-fit and smoothness, where an increase in the smoothing parameter $h$  leads to an increased emphasis on smoothness. We discuss the choice of $h$ in more detail in Section \ref{crossval}. As $h\rightarrow\infty$, the penalty dominates the log-likelihood, leading to a sequence of estimated coefficients $\gamma_{k}$ that lie on a straight line. Thus, we obtain the parametric regression model given in (\ref{linreg}) as a limiting case. Similarly, we can obtain parametric regression models with arbitrary polynomial order $q$ of the predictor as limiting cases by considering $(q+1)$-th order differences in the penalty. The common parametric regression models thus are essentially nested within the class of semiparametric models that we consider.

In general, we might be interested in multiple regressions within a single mark-recapture-recovery model, potentially such that several of these are to be modeled in a nonparametric way. For example, we might want to model both the survival probability and the recapture probability as a nonparametric function of some covariate(s), or model the survival probability separately for different age classes. For illustrations of the latter case, see our real data applications in Section \ref{appl}. Suppose that $M$ different regression functions are to be modeled in a nonparametric way. Then we define the smoothing parameter vector as $\mathbf{h} = (h_1,\ldots,h_M)\in \mathbb{R}_{\geq 0}^{M}$ and consider the global penalty term
 \begin{linenomath*}
\begin{equation*}
 \mathcal{P}(\mathbf{h}) = \sum_{j=1}^M \dfrac{h_j}{2} \sum_{k=3}^K (\Delta^2 \gamma_{j,k})^2\ ,
\end{equation*}
 \end{linenomath*}
where $\gamma_{j,1}, \ldots , \gamma_{j,K}$ are the coefficients used in the representation of the $j$-th regression function, $j=1,\ldots,M$.

\section{Inference}\label{infer}

\subsection{Estimation \& uncertainty quantification}\label{estim}

In each of the scenarios described in Sections \ref{envi}--\ref{inditv}, all model parameters -- including the coefficients $\gamma_{j,k}$ used in the linear combinations of B-splines -- can be estimated simultaneously by numerically maximizing the penalized log-likelihood of the respective model. The numerical maximization is carried out subject to well-known technical issues arising in all optimization problems, including parameter constraints and local maxima of the likelihood. In the scenario with individual-specific time-varying continuous covariates, the number of intervals used in the discretization of the covariate process, $m$, and the considered range of the covariate process need to be chosen. A detailed discussion of how to make these choices is given in \citet{lan13}.

Uncertainty quantification, on both the estimates of parametric parts of the model and on the function estimates, can be performed using a bootstrap \citep{efr93}. A nonparametric bootstrap can be implemented by sampling with replacement from the set of observed encounter histories (typically a stratified sample is used, conditional on the number of initial captures at each capture occasion). Alternatively, when only the summary statistics corresponding to the m- and d-arrays are presented (in the case of only environmental covariates), a parametric bootstrap can be implemented, sampling new histories conditional on the number of new recruits into the dataset for each capture occasion. In all cases, we can obtain pointwise confidence intervals for the estimated regression functions, at specific values of the covariate, as the corresponding quantiles obtained from the bootstrap replications. These pointwise confidence intervals can also be used as a basis for obtaining simultaneous confidence bands for the complete regression functions following an approach proposed in \citet{kri10}. The idea is to rescale the pointwise confidence bands with a constant factor until a certain fraction of complete functions from the set of bootstrap replications is contained in the confidence band. By construction, these simultaneous bands use the pointwise intervals to assess local uncertainty about the estimated function and inflate this local uncertainty such that simultaneous coverage statements are possible.


\subsection{Choice of the smoothing parameter}\label{crossval}

Cross-validation can be applied in order to choose the smoothing parameters in a data-driven way.
If we are dealing with only environmental covariates, and data in the form of m- and d-arrays (see Section \ref{envi}), then it will usually be feasible to conduct a leave-one-out cross-validation. That is, we successively consider each of the $T-1$ rows of the m-array and d-array as a validation sample, with the remaining rows forming a calibration sample. The model is calibrated by estimating the parameters considering only the calibration sample as data. Proper scoring rules \citep{gne07} can then be applied on the validation sample in order to assess the calibrated model for any given smoothing parameter vector $\mathbf{h}$. We consider the (non-penalized) log-likelihood of the validation sample, under the calibrated model, as the score of interest. For each $\mathbf{h}$ considered, the average over the corresponding $T-1$ scores then serves as a (relative) measure of its suitability, such that from some set of possible vectors the $\mathbf{h}$ with the highest average score is selected.

In the case of individual-specific covariates, a leave-one-out cross-validation, successively considering each individual's history as a validation sample, will often be computationally infeasible. Instead, we generate $k$ random partitions of the capture histories such that in each partition a suitable proportion of encounter histories constitutes the calibration sample, e.g.\ 90\%, while the remaining form the validation sample. For each partition, the model is fitted to the calibration sample, and the smoothing parameter vector is then scored with the log-likelihood value of the validation sample. We then consider the average over the $k$ scores as criterion based on which we choose $\mathbf{h}$ from some grid. The number of samples $k$ needs to be high enough to give meaningful scores (i.e.\ such that the scores give a clear pattern), but must not be too high to allow for the approach to be computationally feasible. This cross-validation approach has been successfully applied in similar settings \citep{lan15a,lan15b}.
In both the generalized and in the above leave-one-out cross-validation approach, the optimal $\mathbf{h}$ is chosen from some pre-specified set of possible vectors. 

An alternative, less computer-intensive approach for selecting the smoothing parameters is based on the Akaike Information Criterion (AIC), calculating, for each smoothing parameter vector considered, the associated AIC-type statistic
$$ \text{AIC}_p = -2 \log \mathcal{L} + 2 \nu.  $$
Here $\mathcal{L}$ is the unpenalized likelihood under the given model (fitted via penalized maximum likelihood), and $\nu$ denotes the effective degrees of freedom, defined as the trace of the product of the Fisher information matrix for the unpenalized likelihood and the inverse Fisher information matrix for the penalized likelihood \citep{gra92}. Using the effective degrees of freedom accounts for the effective dimensionality reduction of the parameter space resulting from the penalization. From all smoothing parameter vectors considered, the one with the smallest $\text{AIC}_p$ value is chosen.

\section{A simulation study}\label{simul}

We initially assess the performance of the suggested approach on simulated data. We focus on the scenario with individual-specific time-varying covariates, since it is the most challenging case. We conducted 100 simulation experiments, in each experiment considering simulated encounter histories and covariate values for $N = 600$ individuals, each of them observed on at most $T = 10$ occasions. For each individual, the time of the initial capture occasion was drawn uniformly from $\lbrace 1,\ldots,9 \rbrace$, with the individual's age specified to be 0 at the initial occasion (adding 1 with each occasion). We consider a case with two age classes: age class 1 for age $<2$ (first-years and yearlings) and age class 2 for age $\geq 2$ (adults). The capture and recovery probabilities were considered to be constant and set equal to $0.6$ and $0.4$, respectively. For each individual, its covariate values -- which could for example correspond to the individual's time-dependent weight -- were simulated using an AR(1)-type recursion:
 \begin{linenomath*}
\begin{equation*}
 w_{i,t} = w_{i,t-1} + \eta_{a_{i,t}} \left( \mu_{a_{i,t}} - w_{i,t-1} \right) + \sigma_{a_{i,t}} \varepsilon_{i,t} ,
\end{equation*}
 \end{linenomath*}
where $a_{i,t} \in \lbrace 1,2 \rbrace$ indicates the age class individual $i$ is in at time $t$, and $\varepsilon_{i,t} \overset{iid}{\sim} \mathcal{N}(0,1)$. The covariate values at the initial capture occasions were drawn from a $\mathcal{N}(\mu_0,\sigma_0)$. The parameters of the covariate process were chosen as $(\mu_0,\mu_1,\mu_2) = (-1.4,1,1.3)$, $(\sigma_0,\sigma_1,\sigma_2) = (0.4,0.5,0.4)$ and $(\eta_1,\eta_2) = (0.5,0.8)$. For given covariate values, the survival histories were generated using either of two survival functions --- corresponding to the two different age classes --- so that 
$ \phi_{i,t} = \phi^{(a_{i,t})}(w_{i,t})$, 
where  $\phi^{(1)}(w_{i,t})$ and $\phi^{(2)}(w_{i,t})$ were specified as
 \begin{linenomath*}
\begin{align*}
\phi^{(1)}(w_{i,t}) & =  \begin{cases}
 	\text{logit}^{-1}\bigl(2-0.3(w_{i,t}-0.5)^2\bigr) & \text{ if } w_{i,t} < 0.5;\\
 	\text{logit}^{-1}(2) & \text{ otherwise};
 \end{cases} \\
\phi^{(2)}(w_{i,t}) & =  \text{logit}^{-1}\bigl(\sin (2.5(w_{i,t}+0.8)+0.45)+1.3+0.7 w_{i,t}\bigr).
\end{align*}
 \end{linenomath*}
The function $\phi^{(1)}(w_{i,t})$ mimics a threshold effect, whereas $\phi^{(2)}(w_{i,t})$ exhibits a highly nonlinear pattern.  In each simulation run, the functions $\phi^{(1)}(w_{i,t})$ and $\phi^{(2)}(w_{i,t})$ and all parameters were estimated simultaneously by fitting the semiparametric model to the simulated data (as described in Section \ref{inditv}, with  $\phi^{(1)}(w_{i,t})$ and $\phi^{(2)}(w_{i,t})$ nonparametrically estimated using the strategy described in Section \ref{npreg}). We used $m=50$ intervals in the discretization of the covariate process and $K=15$ basis B-splines in both the representation of $\phi^{(1)}(w_{i,t})$ and in that of $\phi^{(2)}(w_{i,t})$.

We implemented both the generalized cross-validation approach (considering 10 folds) and the AIC-based approach to select smoothing parameters, choosing from the set $\{2^{-2},2^{0},2^{2},2^{4},2^{6},2^{8},2^{10}\}$ for both $\phi^{(1)}(w_{i,t})$ and $\phi^{(2)}(w_{i,t})$, in each simulation run. For both approaches, we estimated the mean integrated squared error (MISE) for the two functional estimators:
$$ \widehat{\text{MISE}}_{\hat{\phi}^{(j)}} = \frac{1}{100}\sum_{z=1}^{100} \left( \int_{w_{\text{min}}}^{w_{\text{max}}} \biggl( \hat{\phi}_{z}^{(j)}(w_{i,t}) - {\phi}^{(j)}(w_{i,t}) \biggr)^2 dw_{i,t} \right)  ,$$
for $j=1,2$, where $\hat{\phi}_{z}^{(j)}(w_{i,t})$ is the functional estimate of ${\phi}^{(j)}(w_{i,t})$ obtained in simulation run $z$. Using cross-validation, we obtained $\widehat{\text{MISE}}_{\hat{\phi}^{(1)}}=0.015$ and $\widehat{\text{MISE}}_{\hat{\phi}^{(2)}}=0.030$, while using the AIC-type criterion, we obtained $\widehat{\text{MISE}}_{\hat{\phi}^{(1)}}=0.014$ and $\widehat{\text{MISE}}_{\hat{\phi}^{(2)}}=0.031$. In the following, we report the estimation results obtained using cross-validation.  

\begin{figure}[!htb]
\includegraphics[width=\textwidth]{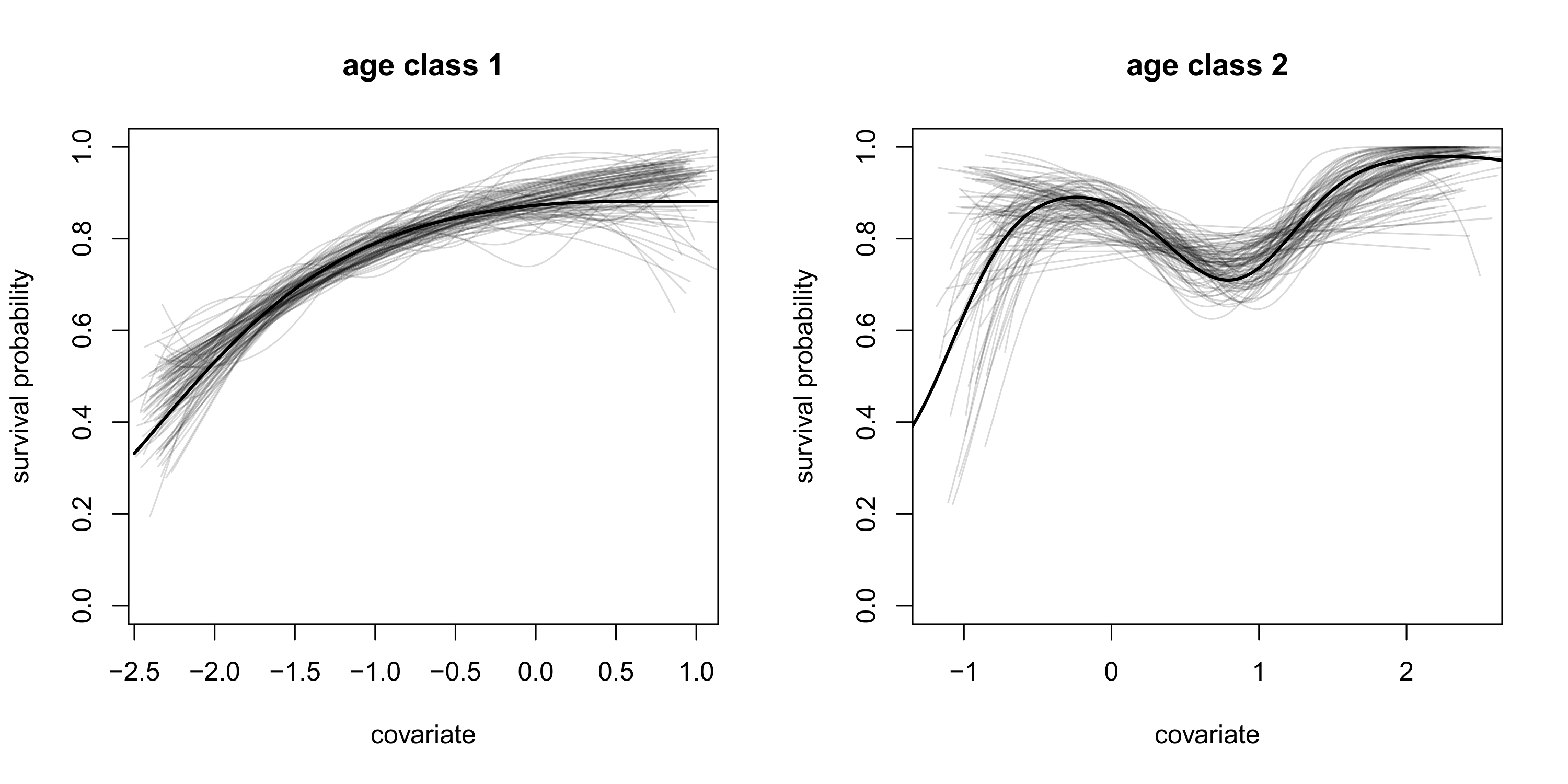}
    \caption{Simulation study: true functions $\phi^{(1)}(w_{i,t})$ and $\phi^{(2)}(w_{i,t})$ considered in the simulation experiments (black lines) and their 100 estimates obtained using the semiparametric modeling approach (gray lines). The range of each estimate was restricted by excluding the first and last 0.005 quantiles of the simulated covariate values.}
    \label{simres}
\end{figure}

Figure \ref{simres} compares the true functions $\phi^{(1)}(w_{i,t})$ and $\phi^{(2)}(w_{i,t})$ that were used to simulate the data, and the estimates obtained using the semiparametric approach. The nonlinearities of the theoretical survival functions were captured fairly well in most of the 100 simulations, with the variance substantially increasing at the boundaries of the range of observed covariate values, as would be expected. However, both functions were clearly oversmoothed in a couple of simulation runs, and undersmoothed in a few others.

\begin{table}[!htb]
\caption{Simulation study: mean relative biases (MRB, e.g.\ $1/100 \sum_{i=1}^{100} (\hat{\lambda}_i-\lambda)/\lambda$, where $\hat{\lambda}_i$ is the estimate obtained in the $i$-th simulation run) and mean standard deviations (MSTD) of the  estimators, based on fitting the model to 100 sets of simulated data.}
\label{simtab}
\begin{center}
	\begin{tabular}{lcc}
		& MRB & MSTD \\
		\hline
		$\hat{\lambda}$ & 0.09 & 0.03\\
		$\hat{p}$ & -0.13 & 0.02\\
		$\hat{\mu}_0$ & 0.20 & 0.02\\
		$\hat{\mu}_1$ & -0.17 & 0.16\\
		$\hat{\mu}_2$ & 1.71 & 0.14\\
		$\hat{\sigma}_0$ & -0.48 & 0.01\\
		$\hat{\sigma}_1$ & -0.67 & 0.02\\
		$\hat{\sigma}_2$ & -0.95 & 0.02\\
		$\hat{\eta}_1$ & -0.11 & 0.04 \\
		$\hat{\eta}_2$ & 0.16 & 0.03 \\
		\hline
	\end{tabular}
	\end{center}
\end{table}

The mean relative biases and mean standard deviations obtained for the parameters $\mu_0$, $\mu_1$, $\mu_2$, $\sigma_0$, $\sigma_1$, $\sigma_2$,  $\eta_1$, $\eta_2$, $p$ and $\lambda$ are given in Table \ref{simtab}. For this (parametric) part of the model, the highest mean relative bias was found to be $1.71\%$ (in case of $\mu_2$).

\section{Application to real data}\label{appl}

\subsection{Heron data}\label{herons}

We first discuss an application that corresponds to the simpler case involving only environmental covariates (see Section \ref{envi}). Note that a similar setting was considered, in a Bayesian framework, by \cite{gim06} and in a frequentist framework by \citet{hug12}. We consider ring-recovery data on grey herons (\textit{Ardea Cinerea}). Each year, nestlings of the species are ringed across Britain. The BTO keeps record of the number of animals tagged, as well as the number of dead herons recovered during the susequent years. We analyse the data on the herons ringed between 1955 and 1997 and recovered between 1956 and 1998, given in the form of a d-array (the data are available from the Biometrics website as supporting material for \citealp{bes02}). Note that there are no live recaptures in this study. Following \cite{bes02}, we consider three different age classes: first-years (age $<$ 1; age class 1), yearlings (age $\in [1,2)$; age class 2) and adults (age $\geq$ 2; age class 3). We are interested in the relationship between the survival probability of herons in a given year and the climatic conditions that year. As an indicator of the latter, we follow \citet{bes02} and consider the annual number of days below freezing. The historical Central England daily temperatures are available from \texttt{www.badc.rl.ac.uk}, and as the covariate, $w_t$, we use the number of frost days in year $t$ as the number of days with average temperature below zero degree Celsius between April of year $t$ and March of year $t+1$ inclusive. Within each of the three age classes, we model the relationship between the number of frost days and annual survival via a logistic nonparametric regression, such that
 \begin{linenomath*}
\begin{equation*}
\text{logit}(\phi_t) = \sum_{k=1}^K \gamma_{a_t,k} B_k(w_t) ,
\end{equation*}
 \end{linenomath*}
where the index $a_t \in \{ 1,2,3 \}$ indicates the age class at time $t$. Note that when using the d-array structure no additional index $i$ indicating individuals is required. Again following \cite{bes02}, the recovery probability, $\lambda_t$, is considered to be variable over the years, modeled via a parametric logistic regression on time with linear predictor. In total, there are $3K+2$ parameters to be estimated.

\begin{figure}[!htb]
 	\includegraphics[width=\textwidth]{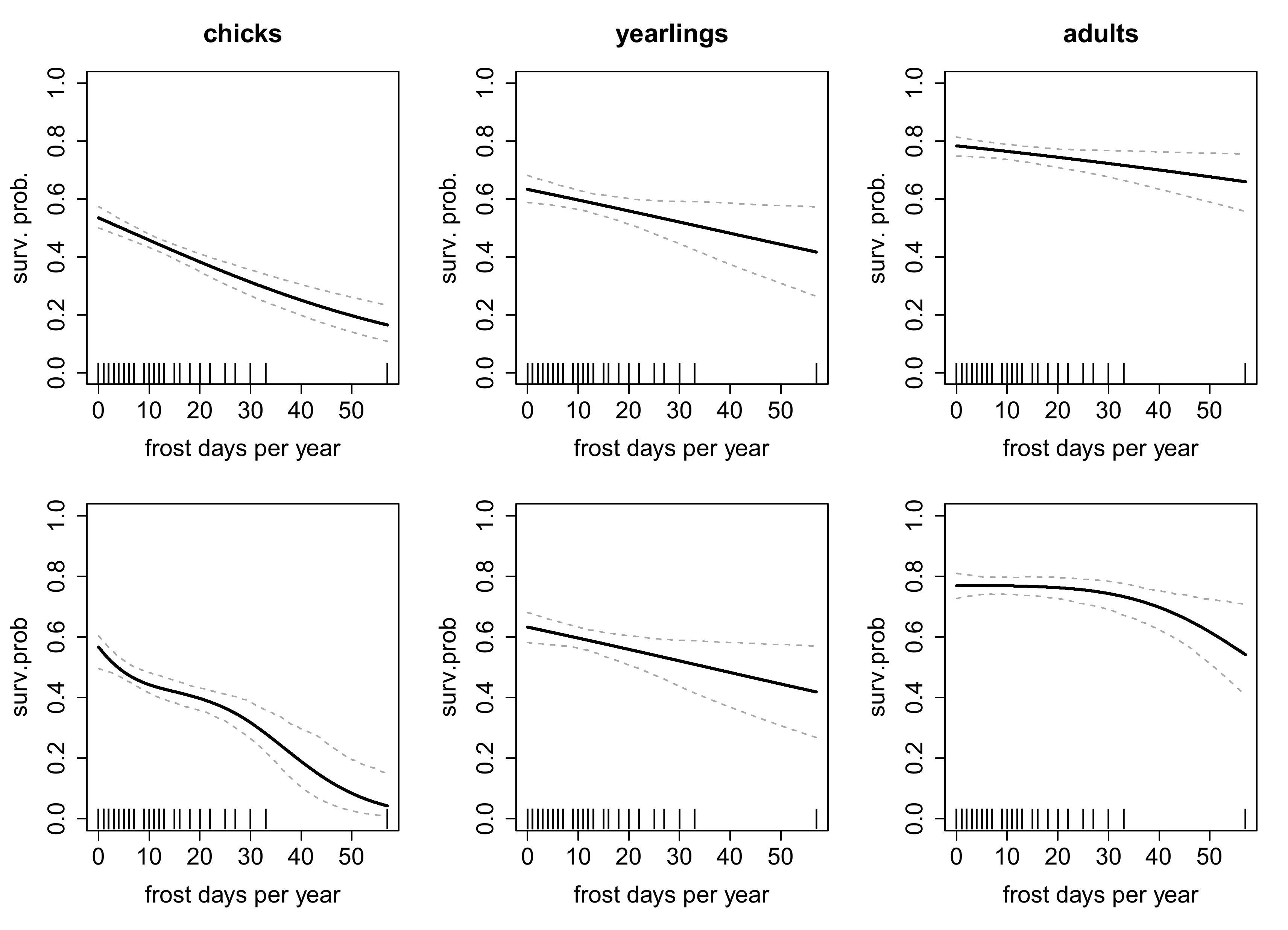} 
    \caption{Heron data: estimated survival probability as a function of the number of frost days per year, according to the parametric (top row) and semiparametric (bottom row) modeling approaches. Solid lines give the maximum likelihood estimates, dashed lines the 95\% pointwise confidence intervals, and vertical dashes indicate the observed values of the covariate.}
    \label{hres}
\end{figure}

Using $K=7$ B-spline basis functions, we conducted a full leave-one-out cross-validation, as described in Section \ref{crossval}, in order to obtain an appropriate vector of smoothing parameters, yielding $\mathbf{h} = (2^{-1},2^{16},2^{2})$. 
Note that the high value of $h_2$ effectively leads to a linear predictor being used in the regression model for the yearlings (cf.\ the plots of the survival functions associated with this age class, given in Figure \ref{hres}). The AIC-based smoothing parameter selection led to $\mathbf{h} = (2^{-1},2^{8},2^{2})$ and virtually identical estimates.

For $\mathbf{h}=(2^{-1},2^{16},2^{2})$, we fitted the semiparametric model to the data, obtaining 95\% pointwise confidence intervals via a parametric bootstrap, as described in Section \ref{estim}, considering 500 samples. On an i5 CPU, at 2.27GHz and with 4GB RAM, fitting the semiparametric model took only a few seconds. For comparison, we repeated the same exercise for the fully parametric (logistic) model. The estimated regression functions and the corresponding 95\% pointwise confidence intervals are displayed in Figure \ref{hres}. The recovery probability was estimated to have slightly declined over time, which is in agreement with the findings reported by \citet{bes02}. 

In both the parametric and the semiparametric analysis, the survival probabilities are consistently lowest for the first-years and highest for the adults, as would be expected, with the youngest birds more susceptible to environmental conditions. In addition, the rate of increase in mortality is largest for first-years and smallest for adults with increasingly harsher environmental conditions (namely, the number of frost days over the winter months). Of particular interest for both first-years and adults is the observed relationship between the survival probability and the very harsh winter in 1962 with an observed environmental covariate value of 57. For the adults, the semiparametric approach suggests that the survival probability is generally constant for the majority of years, but declines for particularly harsh winters, such as that in 1962. This suggests that some form of threshold model may be appropriate (though obviously given the distribution of the observed covariate values it would not be possible to identify the exact threshold without additional data). Within the parametric (logistic) regression model the influence of the particularly harsh winter in 1962 influences the slope of the regression over all years, due to the constrained parametric relationship. A similar though less dramatic misfit of the parametric model may be present for the first-years, where there is some nontrivial structure in the estimated function.

\subsection{Soay sheep data}

Next we discuss an application of the semiparametric approach in the more difficult scenario with an individual-specific and time-varying continuous covariate (as described in Section \ref{inditv}). We consider the capture histories of Soay sheep (\textit{Ovis Aries}) that were born and tagged between 1985 and 2009 on the Island of Hirta (Scotland). 
Each summer, field visits were made that involve, \textit{inter alia}, captures, searches for dead animals and weighings. We consider only female sheep, with at least one recorded weight, leading to a total of 1344 individual capture histories, among which 900 were recovered dead during the observation period. We assume that the survival probability is a function of the individual-specific time-varying weight, $w_{i,t}$, noting that the primary cause of mortality is starvation, with the risk of dying from starvation being highest for young individuals. Following \cite{bon10}, we consider four different age classes, namely lambs (age $<$ 1; age class 1), yearlings (age $\in [1,2)$; age class 2), adults (age $\in [2,7)$; age class 3) and seniors (age $\geq$ 7; age class 4), and model the survival probability as
 \begin{linenomath*}
\begin{equation}\label{agesurv}
\text{logit}(\phi_{i,t})  = \sum_{k=1}^K \gamma_{a_{i,t},k} B_k(w_{i,t}) ,
\end{equation}
 \end{linenomath*}
where the index $a_{i,t} \in \{ 1,2,3,4 \}$ indicates the age class of individual $i$ at time $t$. Thus, four regression functions are estimated, corresponding to the four different age classes. Following \citet{lan13}, we assume the weight of each individual to evolve over time according to the AR(1)-type process
 \begin{linenomath*}
\begin{align*}
w_{i,c_i} & \sim \mathcal{N} (\mu_0,\sigma_0), \\
w_{i,t} & = w_{i,t-1} + \eta_{a_{i,t}} \left( \mu_{a_{i,t}} - w_{i,t-1} \right) + \sigma_{a_{i,t}} \varepsilon_{i,t} \; (\text{for } t>c_i) .
\end{align*}
 \end{linenomath*}
Evidence for the suitability of this process is given in \citet{lan13}. Finally, both the recapture and the recovery probabilities are modeled as time-dependent. To fit this model we simultaneously estimate all $4K+62$ parameters, including the coefficients of the B-splines (4K), the recapture and recovery probabilities (24 each) and the parameters determining the covariate process (14).

We used $K=15$ B-spline basis functions in the representation (\ref{agesurv}) and $m=25$ intervals in the discretization of the covariate process (cf.\ Section \ref{inditv}). We note that the smooth functions estimated using $m=50$ and $100$ (not shown) were visually indistinguishable from those obtained using $m=25$, confirming corresponding analyses in \citet{lan13}. Using cross-validation to simultaneously choose the four different smoothing parameters, associated with the survival functions for the four different age classes, is computationally infeasible in the given scenario. Instead, we ran separate cross-validations for the different age classes, as follows. First, we fitted the fully parametric model described in detail in \citet{lan13}. For age class $a \in \{ 1,2,3,4\}$, we selected the optimal smoothing parameter for the estimation of the corresponding function via cross-validation, as described in Section \ref{crossval}, but regarding all model parameters except the coefficients $\gamma_{a,-K},\ldots,\gamma_{a,K}$ as nuisance parameters. The nuisance parameters were initially fixed at the estimates obtained from the fully parametric model, such that within the cross-validation only the coefficients $\gamma_{a,-K},\ldots,\gamma_{a,K}$ were estimated in the calibration stage. 
This strategy yielded the smoothing parameter vector $\tilde{\mathbf{h}} = (2^{16},2^2,2^{-2},2^{16})$. To further refine the choice of the smoothing parameters, we repeated the same type of cross-validation, only now holding the nuisance parameters fixed at the estimates obtained from the preliminary semiparametric model obtained using the smoothing parameter vector $\tilde{\mathbf{h}}$. This ultimately yielded the smoothing parameter vector $\mathbf{h} = (2^{16},2^2,2^{-1},2^{16})$. The high values of $h_1$ and $h_4$ effectively lead to linear predictors being used in the regression models for the lambs and seniors. In this example, the Fisher information matrix was singular, such that we did not implement the AIC-based smoothing parameter selection.

\begin{figure}[!htb]
\includegraphics[width=\textwidth]{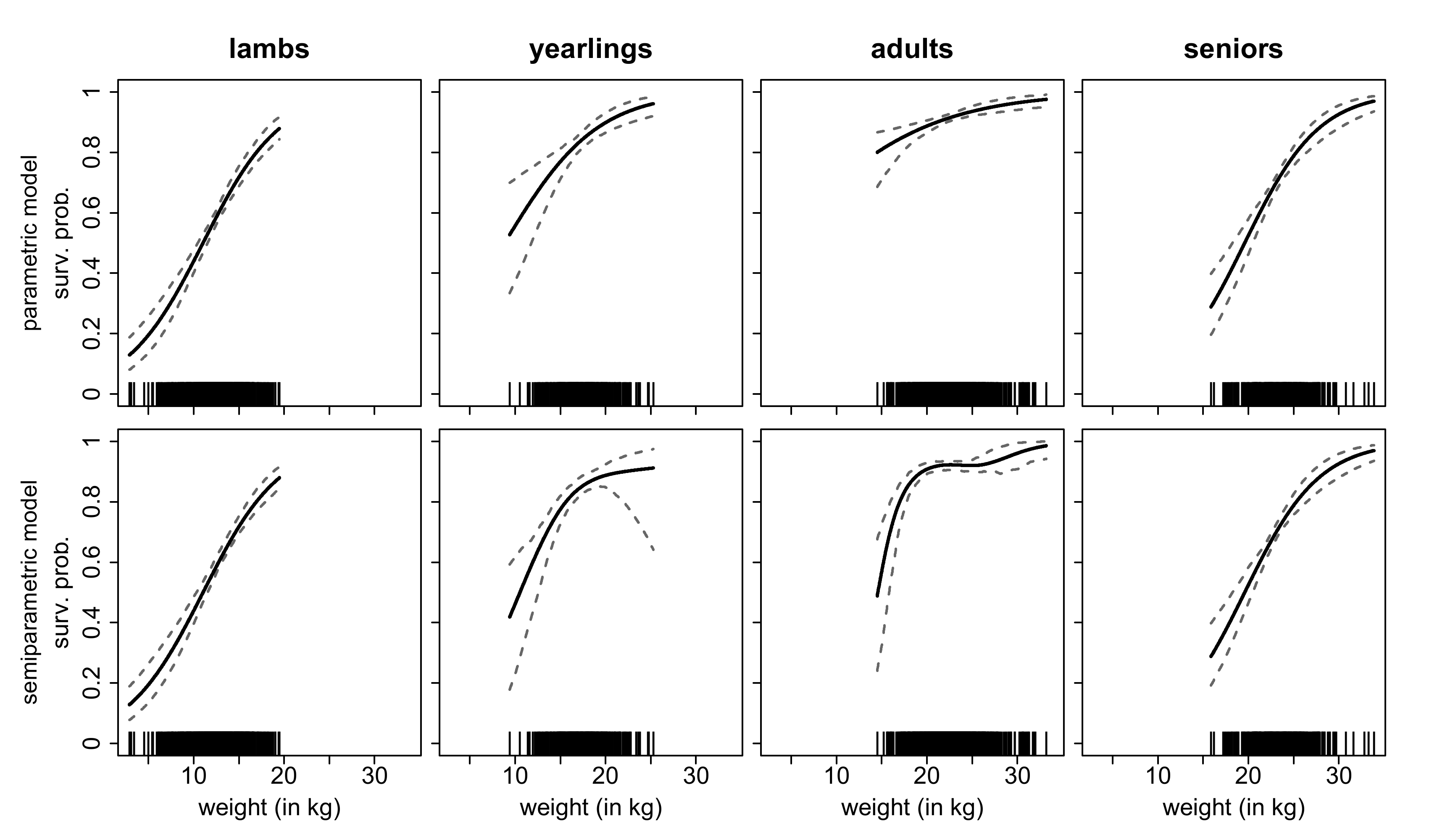}
    \caption{Soay sheep: estimated survival probability as a function of the weight in kilograms, according to the parametric (top row) and semiparametric (bottom row) approaches. Solid lines give the maximum likelihood estimates, dashed lines the 95\% pointwise confidence intervals, and vertical dashes indicate the observed values of the covariate.}
    \label{sres}
\end{figure}

For $\mathbf{h} = (2^{16},2^2,2^{-1},2^{16})$, we fitted the semiparametric model to the data, which took about four hours on an octa-core i7 CPU, at 2.7 GHz and with 4 GB RAM. Note that the substantial decrease in the computational time, compared to the times given in \citet{lan13}, was achieved by writing the main parts of the likelihood calculation in C++. We calculated 95\% pointwise confidence intervals via a nonparametric bootstrap, as described in Section \ref{estim}, considering 200 samples. As for the herons, we repeated the same exercise for the fully parametric model. The estimated regression functions are displayed in Figure \ref{sres}, together with the corresponding 95\% pointwise confidence intervals.

The largest differences between the results obtained from the parametric and semiparametric approaches is found for the adults. In the parametric analysis (here but also in other previous analyses, e.g.\ \citealp{kinbc08}, and \citealp{bon10}, who considered slightly different data), the adult survival probability is alone in not having a significant decrease for individuals of relatively low weight. However, the more flexible semiparametric approach does indicate that there is a sharp decline in the survival probabilities for adults for individuals of relatively low weight, with these individuals having a survival probability that is comparable to yearlings of low weight. Thus, irrespective of age, a relatively low weight, which could be a symptom of poor condition, high parasite load or disease, seems to lead to an increased mortality rate. Furthermore, for yearlings and adults the results obtained from the semiparametric model indicate a minor threshold effect for individuals over some given weight, but there is relatively high uncertainty in case of the yearlings.

\section{Discussion}\label{discuss}

In this manuscript, we have presented a unified inferential framework for semiparametric mark-recapture-recovery models, allowing for any types of covariates of interest. The considered maximum penalized likelihood methods constitute a powerful alternative to the Bayesian approach suggested by \citet{gim06} and \citet{bonts09}. Our work builds on \citet{via10}, \citet{hug12,hug12b}, extending those authors' P-spline-based approaches to the difficult yet important case of individual-specific and stochastically time-varying covariates. The proposed semiparametric modelling approach can be applied more widely to alternative capture-recapture(-recovery)-type models, including for example the conditional trinomial approach \citep{catmt08}, removing the need to model the missing time-varying individual covariate values; stopover models (\cite{pledger09}; removing the conditioning on initial capture) and closed populations (assuming no births, deaths or migrations so that for these models the capture probabilities are often assumed to be a function of covariates).

In the CJS applications to real data on grey herons and on Soay sheep, we demonstrated the relevance of the nonparametric modeling strategy in that it gave us notable new insights into these species' population dynamics. For the herons, the fitted semiparametric model suggests some form of threshold effect in adult survival as driven by environmental condition. We note that previous analyses of heron ring-recovery data, coupled with count data, have suggested a density-dependent threshold for productivity \citep{bes11}. Similarly, for the Soay sheep a threshold model is suggested for adult survival as driven by individual condition, with survival a function of weight only for particularly light-weighted individuals.


In this paper, we focused on the case of a single nonparametric effect within each regression considered. However, the modeling framework also allows for the consideration of additive predictor specifications with multiple nonparametric effects. The same holds true when going beyond simple additive models by including interaction surfaces based on tensor product penalized splines or interaction effects based on varying coefficients. For given smoothing parameters, fitting such models is of a similar complexity as in this paper. However, determining optimal smoothing parameters based on a grid search-type optimization strategy of the cross-validation criterion will become challenging as the number of smoothing parameters increases. In such cases, an interesting alternative would be strategies that allow for a direct estimation of smoothing parameters as an integral part of the numerical optimisation procedures. In the context of penalised spline estimation, mixed model based inference has become quite popular, where the formal equivalence between mixed models and penalised splines is utilised to enable restricted marginal likelihood (REML) estimation of the smoothing parameters \citep{rup03,fah04}. Since REML is equivalent to marginal likelihood estimation, this approach is conceptually relatively straightforward to consider also in our model class, though deriving a corresponding algorithmic solution would require an in-depth investigation of the behaviour of the marginal likelihood and its computational approximation (e.g.\ based on Laplace approximation or similar approaches).

\section*{Acknowledgements}
We would like to thank all the volunteers and members of the Soay sheep project who collected the data from the sheep at St.\ Kilda, and the National Trust for Scotland and the Scottish Natural Heritage who have supported this project. In addition, we would like to thank Tim Coulson for providing the data and for discussions with regard to the data. We would also like to thank the volunteer bird ringers and the BTO for collecting and collating the ring-recovery data.

\renewcommand\refname{References}
\makeatletter
\renewcommand\@biblabel[1]{}

\end{spacing}

\end{document}